# Phonon and structural changes in deformed Bernal stacked bilayer graphene

*Otakar Frank[1,2,*], Milan Bouša,[1,3] Ibtsam Riaz[4], Rashid Jalil[4], Kostya S. Novoselov[4], Georgia Tsoukleri [2,5], John Parthenios[2], Ladislav Kavan[1], Konstantinos Papagelis[2,6,*] and Costas Galiotis[2,6]*

[1] J. Heyrovsky Institute of Physical Chemistry of the AS CR, v.v.i., Prague 8, Czech Republic

[2] Institute of Chemical Engineering and High Temperature Chemical Processes, Foundation of Research and Technology-Hellas (FORTH/ICE-HT), Patras, Greece

[3] Department of Inorganic Chemistry, Faculty of Science, Charles University, Prague 2, Czech Republic

[4] School of Physics and Astronomy, University of Manchester, Manchester, UK

[5] Interdepartmental Programme in Polymer Science and Technology, University of Patras, Patras, Greece

[6] Materials Science Department, University of Patras, Patras, Greece

* Corresponding author: otakar.frank@jh-inst.cas.cz, kpapag@upatras.gr

**Abstract**

We present the first Raman spectroscopic study of Bernal bilayer graphene flakes under uniaxial tension. Apart from a purely mechanical behavior in flake regions where both layers are strained evenly, certain effects stem from inhomogeneous stress distribution across the layers. These phenomena such as the removal of inversion symmetry in bilayer graphene may have important implications in the band-gap engineering providing an alternative route to induce the formation of a band-gap.

Graphene, one atom thick membrane,[1] is the thinnest known elastic material, exhibiting exceptional electronic and mechanical properties.[2] Many applications directly exploiting its Young's modulus of ca. 1 TPa and strength over 160 GPa[3] are envisaged or even already tested, such as mechanical resonators,[4] strain sensors,[5, 6] or graphene-based composites.[7] Bilayer or multilayer graphenes, owing to their distinct electronic band structures have extraordinary potential for next-generation optoelectronics and post-silicon nanoelectronics.[8-10]

According to recent calculations a band-gap opening in monolayer graphene by the sole application of a uniaxial strain is hardly feasible.[11-13] The situation could be more promising when a combination of uniaxial and shear strains would be applied, however, the theory predicts large strains between 12 and 17% to open a band-gap.[14] Recently, significant effort has been put into bilayer graphene, which seems to be more promising in terms of opening a sizable band-gap. Inequality between the two layers caused by different carrier concentrations has already proven successful in opening a measurable band-gap[8, 15, 16] due to a local symmetry breaking.[17] Even though external electric fields have been employed to control the energy gaps in several forms of graphene[8, 10, 15, 16] a simple and practical method of tuning the electronic energy gaps based on pure mechanical deformations is still lacking. For this purpose, since uniaxial stress is not supposed to open up a band-gap at reasonably low strain levels,[18, 19] it is worth considering the inducement of an inhomogeneous strain as a viable tool.[20, 21]

At the moment, there is no detailed experimental study on strained bilayer graphene. Hence it is of utmost importance to understand its response upon mechanical loading. Basic experimental data sets already exist for monolayer graphene[3, 5, 22-29] and thus a direct comparison is available to assess the influence of the second layer and the electronic interaction provided by its presence.

With this respect, Raman spectroscopy is the key diagnostic tool to monitor the number of layers and their changes under external force.[30-32] The G band is a first order Raman mode and corresponds to the in-plane, zone center, doubly degenerate phonon mode with $E_{2g}$ symmetry.[33] The D and 2D modes come from a second-order double resonant process between non equivalent K points in the Brillouin zone (BZ) of graphene, involving two zone–boundary phonons for the 2D and one phonon and a defect for the D band.[34]

In general, tensile strain induces phonon softening (red shift) and compression causes phonon hardening (blue shift). In addition, the G peak splits into two components due to symmetry lowering of the crystal lattice in both cases.[24, 26, 35] The sub-peaks denoted $G^-$ and $G^+$ shift at rates ~30-33 and ~10-14 cm$^{-1}$/%, respectively, both under uniaxial tension[24, 26, 29] and low levels of compression.[24] These shift rates agree well with recent calculations.[13, 26] Additionally, the splitting of the 2D peak has been reported for some graphene samples as well.[23, 29, 36] As it turns out, the origin of this behavior is quite complex and the observed effects strongly depend on the excitation wavelength and the mutual orientation of the graphene lattice, strain direction and incident/scattered light polarization.[23, 29, 36-38] Although the splitting and/or asymmetrical broadening indeed influence the observed behaviur, the 2D peak strain-shift rate has been found by various authors between 45 and

65 cm$^{-1}$/%.[23, 24, 26, 28, 29] A biaxial strain induces shift of the Raman peaks at rates approximately double than those observed in uniaxial tension.[27, 39, 40]

In this work, we have undertaken a detailed Raman study of several samples of bilayer graphene embedded in polymer matrix and compared their behaviour under uniaxial tension with that of monolayer parts of the same flakes. A careful examination along and across the whole flake area was conducted in order to identify the presence of inhomogeneities in the induced strain field due to an uneven stress transfer or slippage either along the polymer/graphene interface or between the two graphene layers. Three excitation wavelengths were used to provide a more coherent picture, especially for the 2D band. Raman features, seen for the first time in mechanically deformed graphene bilayer, are discussed with regards to changes in the local symmetry and possible band-gap opening.

Graphene flakes containing both a mono- and a bilayer part were subjected to tensile strain by means of a cantilever beam assembly[24, 28] and their G and 2D bands were monitored by Raman spectroscopy using 785 nm (1.58 eV), 633 nm (1.96 eV) and 514.5 nm (2.41 eV) excitation. The flakes were either laid bare on a polymer substrate or covered by another polymer layer to minimize a possible slippage during loading.

Figure 1 shows the evolution of the Raman G bands in the embedded mono- and bilayer graphene (Flake F1) under tension using the 785 nm laser excitation. As can be seen, both the bi- and monolayer show the same behavior, i.e. their G bands redshift and split into two components due to the removal of degeneracy of the $E_{2g}$ phonon.[24, 26, 35] In Fig.1, both the mono- and bilayer exhibit the same shift rates of -31.3 and -9.9 cm$^{-1}$/% for G$^-$ and G$^+$, respectively, in accordance with previous experimental results[24, 26, 29] as well as theoretical predictions.[13, 26] The configuration of the respective layers can be seen on a micrograph in Figure 1c. As can be deduced from the image, as well as, from the G$^-$/G$^+$ relative intensities of the respective layers, they belong to the same flake, part of which is composed of a single layer, whereas the other part is overlaid by another layer with Bernal (AB) stacking. Therefore the lattice orientation, which can be calculated from the G$^-$/G$^+$ relative intensities,[23, 26, 29, 35] is the same for the mono- and bilayer, namely ~21° (with respect to the zig-zag direction and strain axis). The indicated polarization dependence of the measured spectra, which shows exactly opposite G$^-$, G$^+$ intensities when the incident light polarization is rotated by 90°, clearly confirms the above. The linearity of the sub-band evolution (Figure 1b), as well as, their shift rates indicate an efficient stress transfer from the embedding polymer to the measured graphene. This issue has been recently found to be of a great importance for a proper analysis of the experimental data.[28, 41, 42] To avoid misinterpretation, we scanned the laser beam over the whole flake area at selected stages of the tensile loading and checked the consistency of the Raman shifts.

The 2D band evolution with strain for the mono- (panels a-c) and bilayer (panels d-f) parts of the Flake F1 are shown in Figure 2. In agreement with recent observations,[23, 29, 36] there is a clear splitting in the monolayer 2D peak (Fig. 2a). Under parallel laser polarization, this spectral feature (Fig 2a, red spectra) can be approximated using two Lorentzian lines with shift rates of -41.5 and -22.4 cm-1/% for the lower (purple) and higher (red) frequency component, respectively, as obtained using a linear fit (Fig. 2b). The faster-shifting (lower frequency) component becomes much more intense with increasing strain under parallel polarization (Fig. 2c), in contrast to the perpendicular polarization, where the two components are of a similar intensity throughout the whole experiment (for more details, see ref. [23]). There seems to be a broad consensus that this splitting is caused by unequivalent K-K' pairs in the deformed Brillouin zone and a differing contribution of the so-called inner or outer processes.[23, 29, 36-38] Furthermore, the amount of splitting depends on the orientation of the sample, laser polarization and excitation wavelength.[23, 38] The Raman 2D mode in bilayer graphene (Fig. 2d-f) is expected to be affected by the strain-induced electronic changes since it arises from a double resonance process that involves transitions among various electronic states.[31, 32, 43] Consequently, the larger FWHM of the 2D peak in bilayer graphene compared to that of the monolayer makes the assignment of the deconvoluted peaks a rather difficult task. More specifically, in bilayer graphene having Bernal configuration the dispersion of $\pi$-electrons and phonon bands near the K point of the Brillouin zone split both into two components with specific symmetries (Fig. 2g). The electronic branches split into two conduction ($\pi_1^*$, $\pi_2^*$) and two valence ($\pi_1$, $\pi_2$) bands, while the dispersion is parabolic near the K point (Fig. 2g). The $\pi_1$ and $\pi_1^*$ bands are degenerate at K and the $\pi_2$ and $\pi_2^*$ ones exhibit an energy gap of about 0.8 eV.[32, 43, 44] Group theory for bilayer graphene predicts four dinstict DR processes along the Γ–K–M–K'–Γ direction.[32] It should be noted that along the aferomentioned BZ direction the elecron-phonon coupling between the iTO (in-plane transverse optical) phonons and $\pi$-electrons near K and K´ has the highest oscillation strength.[43] The four processes can be assigned as $D_{ij}$ where i (j) denotes an electron scattered from (to) each conduction band $\pi_{i(j)}^*$ conduction band. The lower ($\pi_1^*$) and upper ($\pi_2^*$) conduction bands belong to different irreducible representations ($T_1$ and $T_2$ respectively). Also, the iTO phonon branch splits into two branches related to the symmetric (assigned as S and correspond to $T_1$ irreducible representation) and antisymmetric (assigned as AS and correspond to $T_2$ irreducible representation) phonons, with respect to the inversion symmetry. According to electron-phonon selection rules the S phonons are connected with the $D_{11}$ and $D_{22}$ processes involving electrons with the same symmetry whereas the AS phonons occur for processes $D_{12}$ $D_{21}$ involving electrons with different symmetries. Since the iTO phonon frequency increases with increasing *q* (due to the Kohn anomaly at K point[32]) the highest (lowest) frequency peak of the 2D band is associated with $D_{11}$ ($D_{22}$) process. Experimentally, the 2D band lineshape of a Bernal-stacked bilayer graphene is fitted by four Lorentzian components, each having the same FWHM of approx. 24 cm$^{-1}$ (Fig. 4d, bottom spectrum).[31, 32]

Under tension, the shape of the bilayer 2D band clearly changes (Fig. 2d). The evolution of the four sub-bands can be monitored credibly only when the fitting is carried out using components of equal widths at a given strain level. Also, in a very recent high pressure doping Raman study of bilayer graphene, Nicolle et al. followed a similar fitting protocol.[45] As can be seen in Fig. 2e, the positions of the components follow linear trends with similar rates of ~ -50 cm$^{-1}$/% for the three lower frequency components ($2D_{22}$, $2D_{21}$ and $2D_{12}$) and a smaller shift rate of -29 cm$^{-1}$/% for the highest frequency component $2D_{11}$. The intensities of the components are changing unevenly too (Fig. 2f); the intensity of the $2D_{12}$ component increases, mainly at the cost of the neighboring $2D_{21}$ and $2D_{11}$ with the latter almost diminishing at higher strains.

It is interesting to compare the two sub-bands of the monolayer 2D peak (Figs. 2a-c) and the two highest frequency components $2D_{11}$ and $2D_{12}$ in the bilayer (Figs. 2d-f). As mentioned above, the $2D_{11}$ component originates from the processes associated with the "original" $\pi_1$ and $\pi_1^*$ bands degenerated at the K point, while the other three lower wavenumber components involve at least one electron or hole from the $\pi_2$ and $\pi_2^*$ bands.[31, 32, 43, 44] Bearing that in mind, a plausible scenario might be suggested to explain the observed shift rates and the evolution of intensities. Let us consider a splitting of the $2D_{11}$ component in the same way as in the monolayer 2D band. Then, what we observe as the $2D_{11}$ component in the strained bilayer – with smaller shift rate and decreasing intensity - would be actually only the higher wavenumber sub-component (compare red symbols in Figs. 2e-f and in Figs. 2b-c). Whereas the lower wavenumber sub-component would gradually shift towards the $2D_{12}$ component and merge with it (blue symbols in Figs. 2e-f vs. purple symbols in Figs. 2b-c), causing its apparent intensity to increase upon stretching. The change of the bilayer 2D band shape recorded under perpendicular laser polarization might be connected with the observed concomitant alterations in the monolayer (black curves in Figs. 2a and d). The discussed $2D_{11}$ splitting can further supported by the work of Mafra et al.[46], who pointed out the importance of the inner scattering processes in Bernal bilayer graphene. Hence the complex line-shape of the 2D Raman signal comes from the contribution of the inner and outer processes due to the anisotropy of both the electron and phonon dispersions.[46] Our experimental data show that the $2D_{11}$ component is more sensitive to modifications in the electronic and/or phononic structure, while the degree of splitting is not that pronounced for the other components. However, more theoretical work about the double resonance mechanisms in graphene systems is necessary in order to understand fully the effect of strain upon the changes to the lineshape of the 2D peak.

As mentioned above, the monolayer 2D peak splitting depends strongly on the excitation wavelength. Therefore, we have examined another flake (F2) composed of mono- and bilayer graphene using a different excitation line of 633 nm (Figure 3). At a certain strain level, which is apparently higher in the monolayer than in the bilayer, a failure in the stress uptake can be seen (labeled with a dotted vertical line in Fig. 3b and d ). Up to the failure, the monolayer 2D peak exhibits no obvious splitting (shift rate of -48.5 $cm^{-1}$/%), but only a slight asymmetry and broadening in the final stages of the experiment. An important issue arises here; Huang et al.[36] and Yoon et al.[29] using the 532 nm (2.33 eV) and 514.5 nm (2.41 eV) excitations, respectively, reported a clear 2D splitting for strain applied essentially along the armchair and zigzag directions. Our measurements,[28] using the 514.5 nm showed a relatively small 2D-peak broadening at a rate of 11 $cm^{-1}$/% and, apparently, absence of splitting. On the contrary, using the 785 nm (1.58 eV) excitation a significant non-linear FWHM enhancement is observed reaching approximately 43 $cm^{-1}$ at strain of only 0.7% for approx. 20° lattice orientation.[23] Furthermore, the data presented here using an 633 nm (1.96 eV) excitation showed a shift rate of 17$cm^{-1}$/% for the FWHM of the 2D (15° lattice orientation), indicating a progressive FWHM increase when the excitation energy is red-shifted. This experimental evidence is in contrast with recent calculations of Venezuela et al.[38] who predicted the opposite trend, namely that the 2D line becomes broader in an assymetric double peak structure, at higher excitation energies up to ultra-violet range. Therefore, our observations confirm that Raman data using various excitation wavelengths and sample orientations are extremely important to fully resolve the double resonance process in graphene systems.

In contrast to the flake F1 (Fig. 2d-f; $\lambda_{exc}$=785 nm), all four Lorentzian components of the bilayer 2D band in flake F2 (Fig. 3c-e; $\lambda_{exc}$=633 nm) exhibit almost identical shift rates of ~ -55 $cm^{-1}$/% and only minor changes in relative intensities. Here, the absence of splitting in the monolayer seems to be reflected in the uniform evolution of the bilayer 2D band components. This justifies our previous statement on the higher sensitivity of the $2D_{11}$ component, which apparently correlates with the 2D band behavior of the neighboring monolayer.

In Fig. 4a, a detailed Raman mapping of flake F2 strained to 0.74% (Fig. 4a; $\lambda_{exc}$=633 nm) reveals inhomogeneities in the stress field across the measured sample. Regardless of the number of layers, the stress uptake is zero at the edges of the flake. Similar behavior has been documented for the monolayer graphene[41] and it indicates that at least up to a certain degree, the stress is transferred to the graphene by a shear mechanism at the interface.

Another local feature depicted by the vertical black line on the monolayer part, which marks a very narrow region with a zero strain, is probably connected with a crack formation in the graphene flake (for additional data, see Supporting Information S1).

As shown in Fig. 4a, there is an obvious difference in the stress uptake between the mono- and bilayer parts in our sample, and this was also reflected (Figs. 3b and d) in the observed differences at the onset of failure for the mono- and bilayer graphenes. In fact, since the bilayer has twice the thickness of the monolayer, it would require a much higher interface shear stress for the same level of axial strain transferred to the flakes. If the ceiling of the interface shear has been reached at 0.5% strain for the bilayer/PMMA system then interface failure (or slippage) is quite likely to occur. This would certainly lead to the relaxation of the stress transferred to the flakes exactly as observed in Figs. 3b and d.

We now turn our attention to local failure events throughout the large flake areas. In Fig.4a, the black ellipse marks a small region on the bilayer not larger than 5 μm close to the bilayer/monolayer boundary, where dramatic spectral changes are observed for the bilayer 2D band. This is in contrast to the rest of the specimen which exhibits the lineshape common to the four peaks of a Bernal bilayer (bottom in Fig. 4b) originating from the different electron-phonon scattering processes discussed above. The Raman signal from the marked region shows a 2D peak completely lacking the characteristic features of an AB stacked bilayer (top in Fig. 4b) and has a lineshape more typical of a monolayer or non-AB stacked bilayer (for other maps displaying additional information from the 2D band deconvolution, see supporting Information S2).[47] This might be due to cohesive (interplanar) shear failure in the bilayer due to the high normal stresses developing in that region due to the possible presence of atomic defects. The possibility of detecting cohesive failure in multi-layer graphenes through the Bernal to non-Bernal transition opens the way for assessing failure at the nanoscale with the use of Raman spectroscopy.

The G band in the Raman spectra from this region of the sample shows a peculiar behavior as well (Fig.4c, top spectrum). In this flake, the prevalent G band shape of the strained bilayer (and the monolayer too) consists of $G^-$ and $G^+$ sub-bands with a $G^-/G^+$ intensity ratio of ~0.65, centered on frequencies corresponding to the level of applied uniaxial deformation (Fig. 4c, bottom).[23, 26, 29] However, in the labeled region, the G band exhibits one intense component (further noted as $G_1$), redshifted to 1578 cm$^{-1}$ from the zero strain position (~ 1584 cm$^{-1}$), but also another weaker band (noted as $G_2$) at 1594 cm$^{-1}$, i.e. at a much higher wavenumber than could be expected taking into account the typical G$^-$, G$^+$ shift values.[24, 26] The assignment of the $G_1$ and $G_2$ bands to the individual layers is highly improbable, since this would have to be accompanied by two clearly separated 2D bands (by approx. 30 cm$^{-1}$). It also has to be noted that the presence of this weak band is strictly limited to the marked region and is always accompanied by the distorted 2D peak shape and a complete absence of the D band, which rules out the assignment of the $G_2$ band to the D' band (see Fig. S3 and related discussion in Supporting Information). This observation might suggest that the two graphene layers experience unequal strain fields, resulting in the inversion symmetry breaking of the bilayer lattice. In bilayer graphene the doubly degenerate $E_{2g}$ branch of the monolayer evolves, at the Γ point, into two doubly degenerate branches $E_g$ and $E_u$ (insets in Fig. 4c).[48] The antisymmetric $E_u$ mode is under normal conditions Raman inactive and can be observed using infrared spectroscopy. The $E_g$ and $E_u$ modes were reported to split due to a weak interlayer coupling in gated[49-51] or chemically doped[52] bilayer graphene. It is tempting to assign the $G_1$ and $G_2$ band to $E_g$ and $E_u$ phonons, respectively. As mentioned above, the redshift of the $E_g$ (~$G_1$) band is about 6 cm$^{-1}$ from the zero strain position. Considering the ambient position of the $E_u$ mode close to 1600 cm$^{-1}$,[53] a similar shift can be ascribed to the $G_2$ band. The much weaker intensity of the $E_u$ mode suggests that the degree of inversion-symmetry breaking is rather low, indicating only a mild mixing between the $E_g$ and $E_u$ modes.[54]

The peculiar G and 2D band features point to a local asymmetry between the two layers of the otherwise Bernal stacked bilayer graphene. This is in accordance with the recent theoretical work of Choi et al.[20], who stated that transverse electric fields across the two layers can be generated without any external electronic sources, thereby opening an energy gap. With this respect, it should be stressed that even though the applied stress was uniaxial and very low (0.74%), our observation shows a relative simplicity of tuning the non-uniform strain fields in the two layers, which is perfectly in line with very recent calculations of Verberck et al.[55]

In summary, we present a systematic uniaxial deformation Raman study of several bilayer graphene samples (and monolayers being parts of the same flakes) embedded in a polymer matrix, using laser energies from the visible to the near-IR range. Our experimental data show that strain directly influences the double resonance bands, while the $2D_{11}$ component is more sensitive to the induced deformations. This work contributes an experimental insight to the various scenarios[23, 29, 36, 38] presented in the recent literature regarding the relative contribution of the inner vis-à-vis the outer process to the 2D Raman peak.

In terms of the mechanical stability, we observed that the interface failure or slippage of the bilayer occurs at lower tension levels compared to the monolayer part of the same flake. Additionally, the Bernal-stacked two layers fully embedded in a matrix are locally susceptible to non-uniform strain field components, which induce a breaking of the bilayer inversion symmetry. This in turn leads to the activation of the infrared $E_u$ mode and the appearance of a single broad 2D band component. The results can be explained considering the band-gap opening, as proposed by recent theoretical predictions.[20, 55] Further work towards a control, at a larger scale, of the strains applied independently to each of the two layers is needed to confirm the viability and the potential of such a mechanism.

ACKNOWLEDGMENT

FORTH / ICE-HT acknowledge financial support from the *John S. Latsis* Public Benefit Foundation (Greece). Part of this work has been funded by 'Graphene Centre' of Foundation of Research and Technology Hellas. K.S.N. is grateful to the Royal Society and European Research Council (grant 207354 – "Graphene") for support. O.F., M.B. and L.K. further acknowledge the financial support of Czech Ministry of Education Youth and Sports (contract No. LC-510), the Academy of Sciences of the Czech Republic (contracts IAA 400400804 and KAN 200100801) and the EC 7th FP project Molesol (256617). Finally, this research has been co-financed by the ESF and Greek national funds through the Operational Program "Education and Lifelong Learning" of the NSRF - Research Funding Program: Heracleitus II.

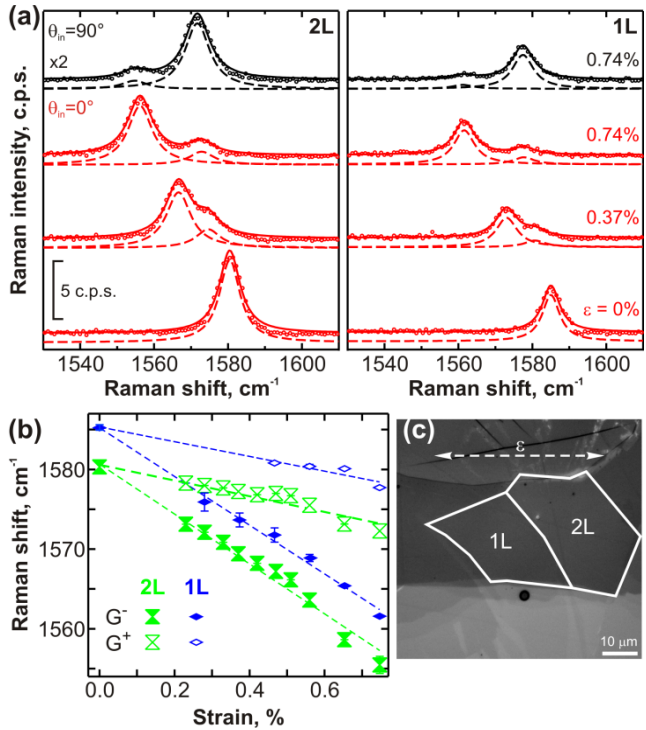

**Figure 1.** (a) G band Raman spectra of strained mono- (right) and bilayer (left) graphene flake F1 excited by 785 nm. Dashed curves are Lorentzian fits of the individual components with solid lines as their convolution and the points represent experimental spectra. Plot (b) shows the evolution of the position of the $G^-$ and $G^+$ sub-bands with strain. Dashed lines represent linear fits to the experimental data. Error bars are one standard deviation to all data measured at the given point and strain level. Panel (c) shows an optical micrograph of the flake F1. The strain axis, $\varepsilon$, is depicted with a horizontal arrow.

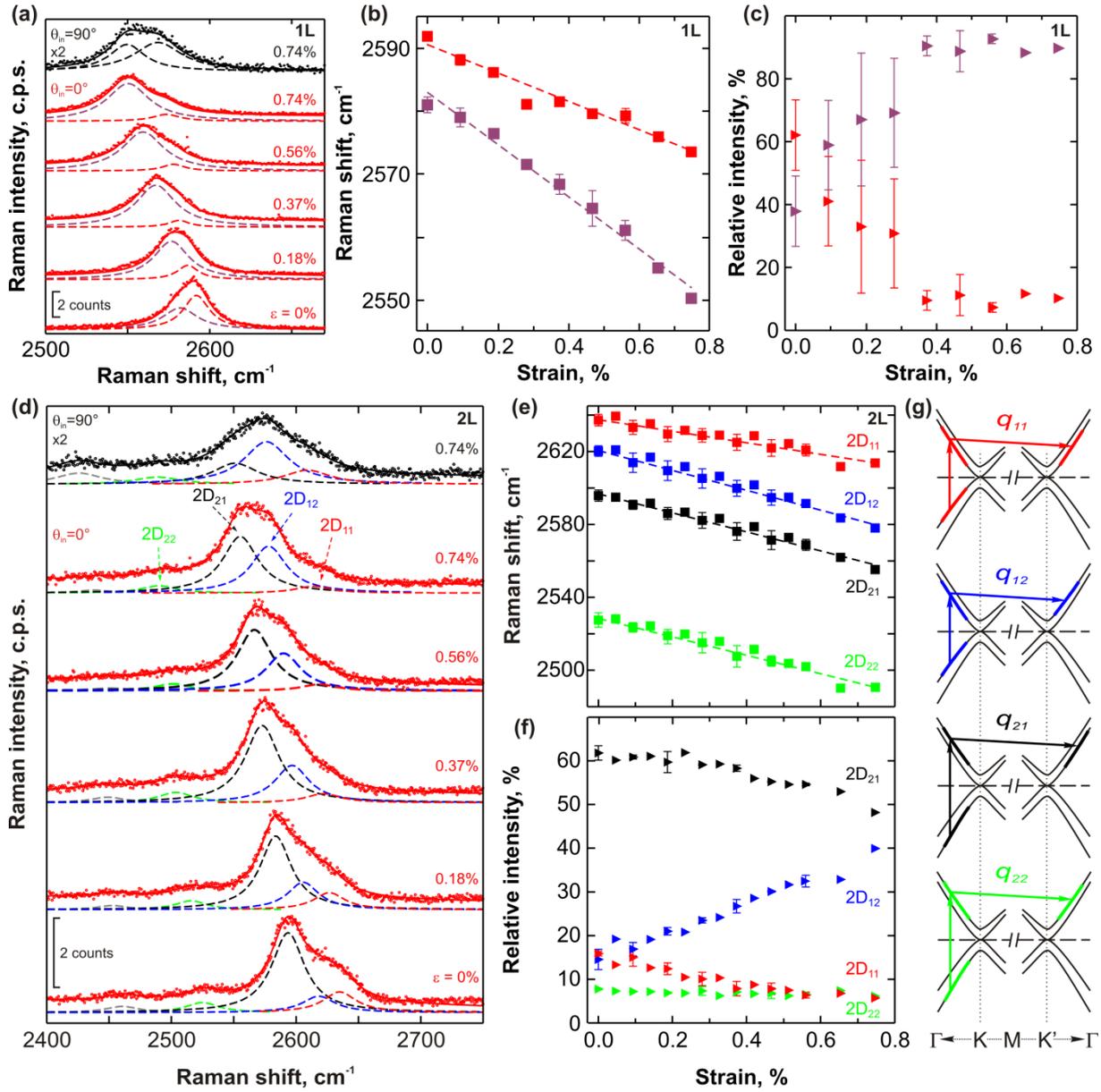

**Figure 2.** 2D band Raman spectra of strained (a) mono- and (d) bilayer graphene flake F1 excited by 785 nm. Dashed curves are Lorentzian fits of the individual components with solid lines as their convolution and the points represent experimental spectra. Plots (b) and (e) show the evolution of the position of the individual Lorentzian components with strain in the mono- and bilayer, respectively, while plots (c) and (f) show the evolution of the relative intensities of the respective components (all with $\theta_{in}=0°$). Error bars are one standard deviation to all data measured at the given point and strain level. (g) Schematic sketches depicting the four processes in bilayer graphene, for a better clarity, only the first scattering $q_{ij}$ events plotted.

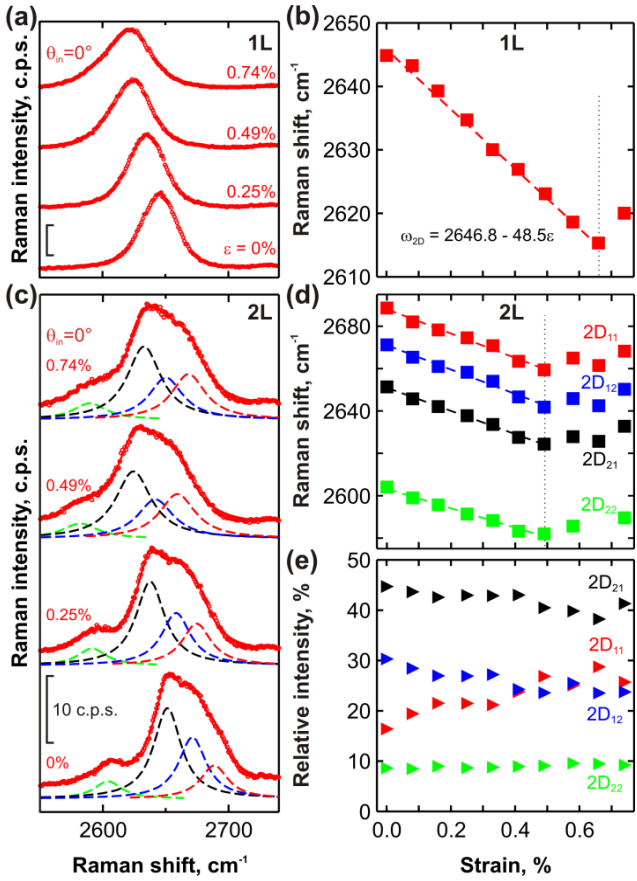

**Figure 3.** 2D band Raman spectra of strained (a) mono- and (c) bilayer graphene flake F2 excited by 633 nm. Dashed curves are Lorentzian fits of the individual components with solid lines as their convolution and the points represent experimental spectra. Plots (b) and (d) show the evolution of the position of the individual Lorentzian components with strain in the mono- and bilayer, respectively, plot (e) shows the evolution of the relative intensities of the respective components for the bilayer part. In panels b and d, dotted lines mark the failure strain level, dashed lines represent linear fits to the experimental data up to the failure.

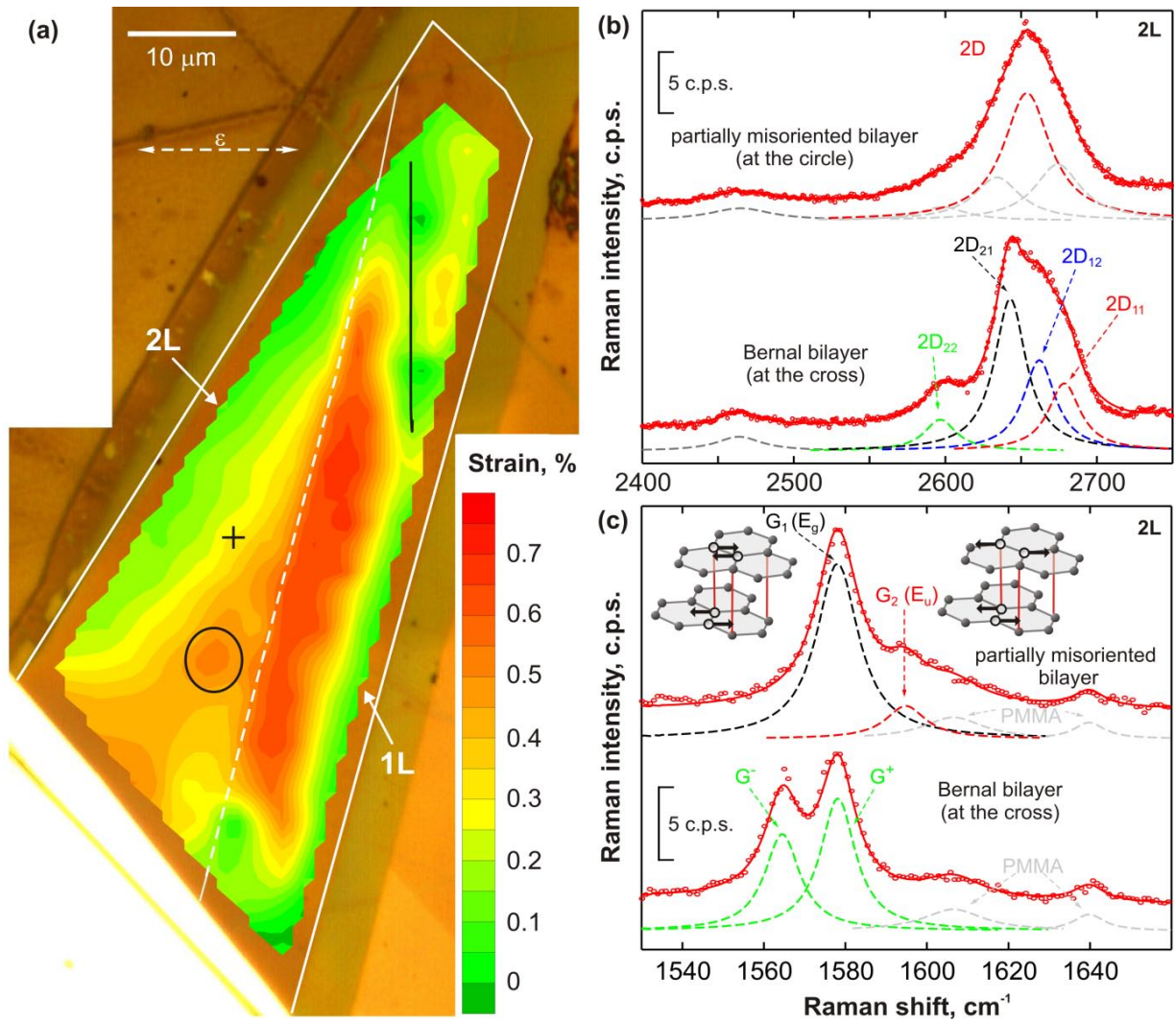

**Figure 4.** (a) Strain map of the flake F2 as calculated from Raman 2D band positions at a nominal tension of 0.74% (see Methods). Example spectra of the 2D (chart b) and G band regions (chart c) of the bilayer part measured inside the black circle and at the cross. Dashed curves are Lorentzian fits of the individual components with solid lines as their convolution and the points represent original spectra.

## Supporting Information

### Materials and Methods

Graphene monolayers were prepared by mechanical cleavage from natural graphite (Nacional de Grafite) and transferred onto the PMMA (polymethylmethacrylate) cantilever beam covered by a ~200 nm thick layer of SU8 photoresist (MicroChem). After placing the graphene samples, either a thin layer of S1805 photoresist (Shipley) or PMMA (MicroChem) was spin-coated on the top. The top surface of the beam can be subjected to a gradient of applied strain by flexing the beam by means of an adjustable screw positioned at a distance $L$ from the fixed end. The deflection $\delta$ was measured accurately using a dial gauge micrometer attached to the top surface of the beam. Furthermore, the total thickness of the beam $t$ and the flake's distance from the fixed end $x$ are taken into account for the calculation of the strain level.

MicroRaman (LabRAM HR, Horiba Jobin-Yvon, France, or InVia Reflex, Rensihaw, UK) spectra were recorded with 785 nm (1.58eV), 633 nm (1.96 eV) or 514.5 nm (2.41 eV) excitations, while the laser power was below 0.9 mW. A 100x objective with numerical aperture of 0.9 is used, and the spot size is estimated to be ~1x2 µm$^2$ for 785 nm excitation and ~1x1 µm$^2$ or less for 633 and 514 nm excitations. The polarization of the incident light was either parallel ($\theta_{in}=0°$) or perpendicular ($\theta_{in}=90°$) to the applied strain axis, while the scattered light polarization was selected, in all cases, parallel to the strain axis ($\theta_{out}=0°$). All peaks in the Raman spectra were fitted with Lorentzians. The background peaks from the embedding polymers were either subtracted after normalization or included in the fitting procedure with fixed positions and widths determined from a separate measurements. The strain in Figure 4 was calculated from the 2D band of the respective 1L and 2L parts: first, the shift rate for every component (or a single peak for the monolayer) was determined from the linear part of the tension experiment before failure, followed by a calculation of the strain for each of the 2D components in bilayer or the single band in monolayer for every spectrum in the map. Finally, in the case of the bilayer, the strain value was determined as an average weighted by the areas of the four components.

### Crack Mapping

As mentioned in the main text, the local feature depicted by the vertical black line on the monolayer part in Fig.4a marks a very narrow region with a zero strain, perpendicular to the strain direction. A closer inspection of the spectra recorded in this region shows a clear presence of the D band (Fig. S1), which is otherwise completely absent in the measured data (even at the edges). In order to confirm the nature of the band, 785 nm excitation has been used as well (fig. S1, bottom), and indeed, the position of the band changed owing to its dispersion behavior. Additionally, a detailed mapping of this particular region has been conducted, which showed the D band appearance is really confined into a narrow strip, not wider than the laser spot size (~ 1 μm).

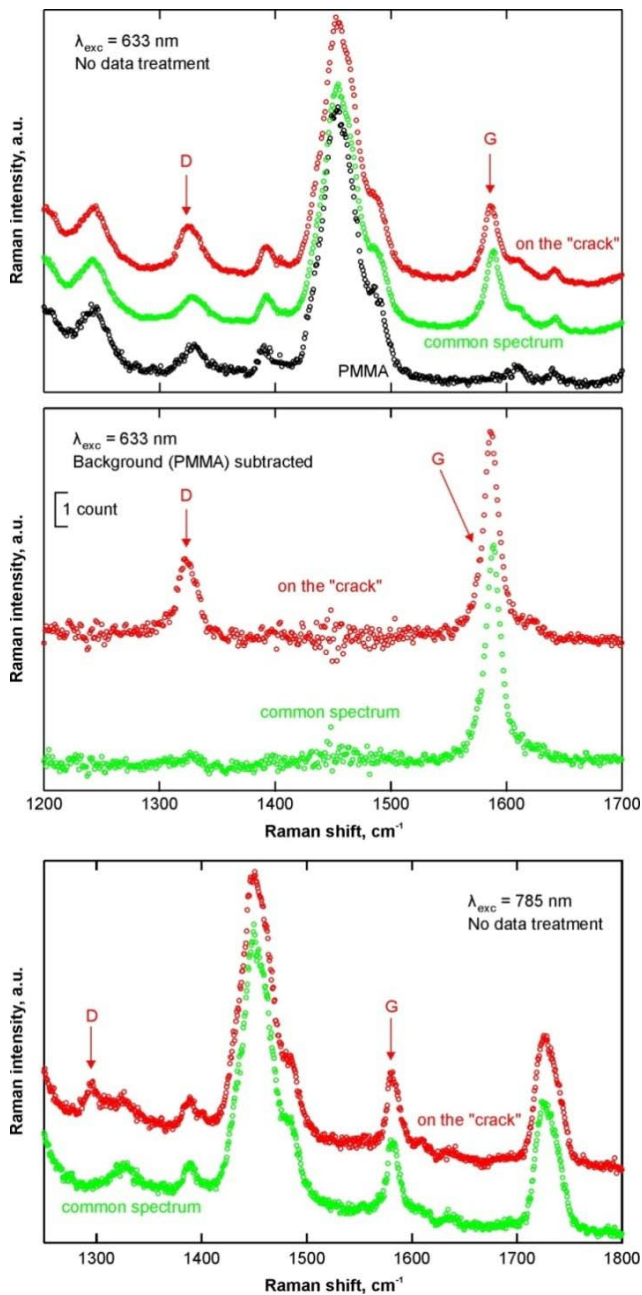

**Figure S1**. Raman spectra measured in the region marked in Fig.4a (main text) with a black vertical line compared to common spectra measured on the flake at zero strain.

At this moment, we might only speculate about the origin of this feature, however given the simultaneous presence of the D band and the zero strain derived from the G and 2D bands position, a reasonable explanation might be a crack present in the graphene itself. Somewhat similar fragmentation of repeatedly strained graphene has been shown by Young *et al.*[1], who assumed its origin in the loss of stress transfer due to cracks in the embedding polymer. We have observed no such cracks in the top PMMA layer, however, their presence in the bottom SU8 cannot be ruled out. Since it is difficult to assume alone the strain in the order of 0.74% would be responsible for fracturing the graphene, an edge from the ruptured polymer might locally damage the monolayer.

**Additional Raman maps; complementary to Figure 4a**

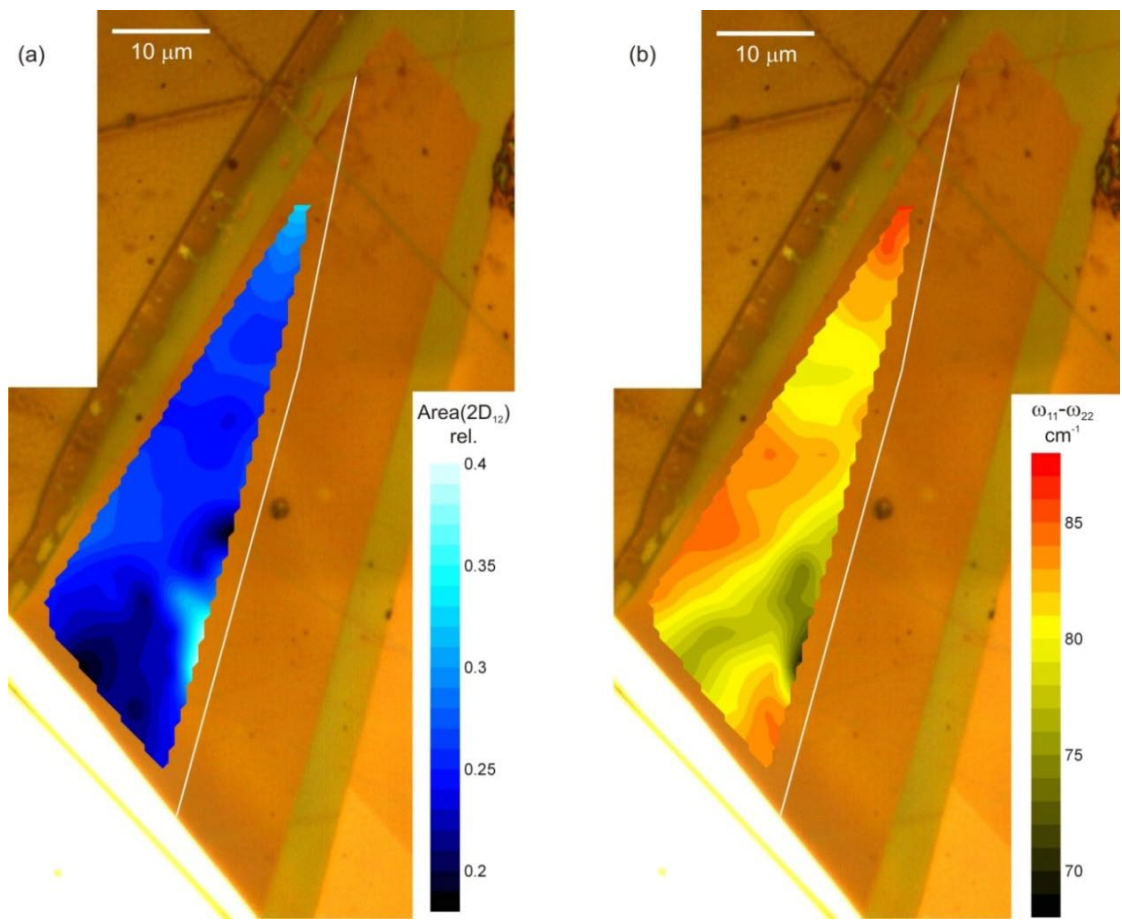

**Figure S2.** Raman maps of (a) the relative area of the $2D_{12}$ component of the bilayer and (b) the difference in position of the $2D_{11}$ and $2D_{22}$ components.

**Discussion on the origin of the $G_1$ and $G_2$ bands**

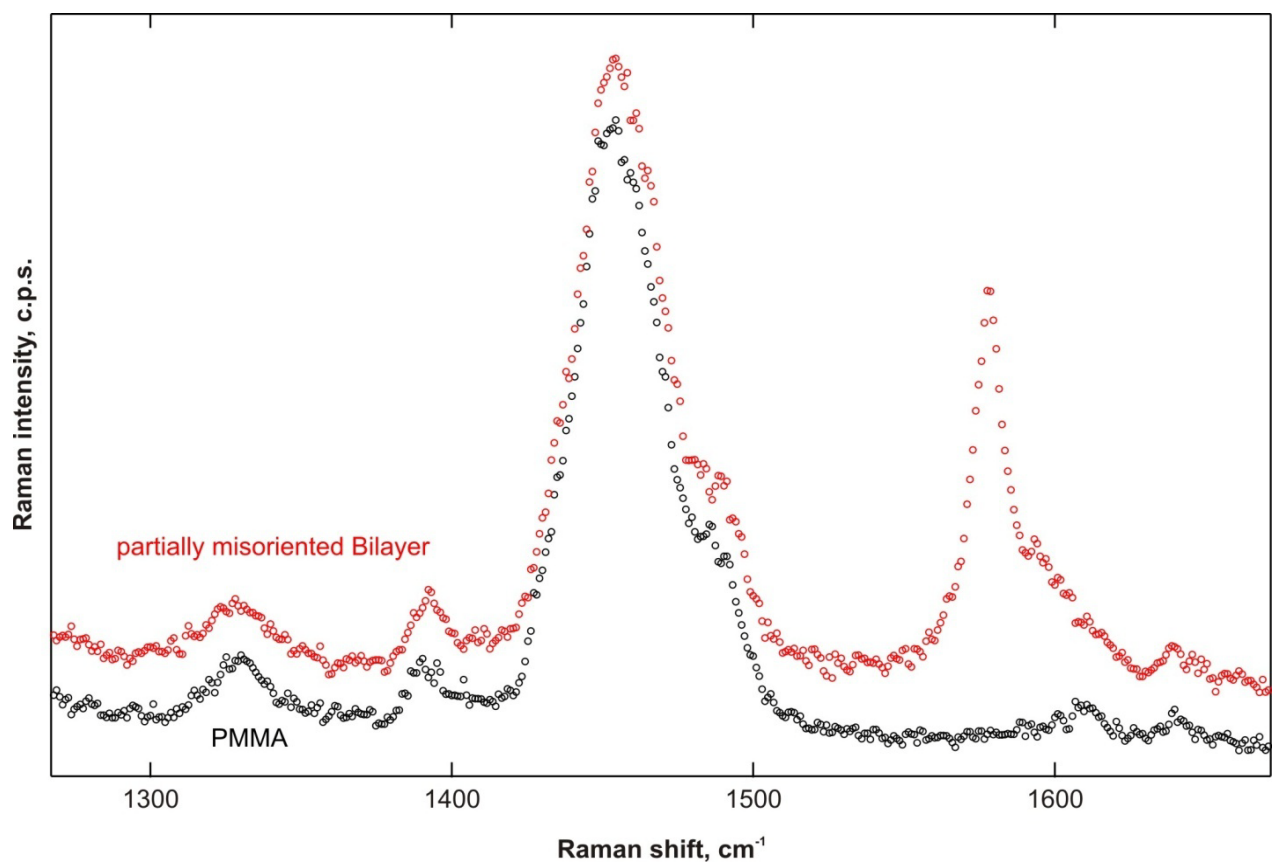

**Figure S3.** Raman spectrum of the partially misoriented Bernal bilayer graphene (from Fig. 4c – top, main text) with the range extended down to 1250 cm$^{-1}$ and comparison to the spectrum of the PMMA substrate.

Several other explanations of the origin of the $G_1$ and $G_2$ bands from the Figure 4c (main text) have been considered; none of them turns out to be valid enough. The possibility of the $G_2$ band being in fact the D' (the intra-valley defect-induced mode) has been ruled out because of the absence of the D band in the particular spectra (see Fig. S3 and compare to the D band presence as shown in Fig. S1). The D' band in carbonaceous materials appears never alone – without the presence of the D band – and it is also less intense. Given the intensity of the $G_2$ band, the intensity of the D band would have to be at least comparable to that presented in Fig. S2. Conversely, there is no such shoulder as the $G_2$ band visible in the spectra containing the D band in Fig. S1.

The hypothesis of the $G_1$ and $G_2$ bands each reflecting a totally opposite strain state of the two layers – tension and compression, respectively – can be neglected too. In such a case, not only we should see a further splitting of each of the bands (into G$^-$, G$^+$), but mainly the 2D band would have to exhibit two distinct single-Lorentzian components, separated by approximately the double the difference between the $G_1$ and $G_2$, i.e. 32 cm$^{-1}$. Similarly, the possibility of a simultaneous sampling of the mono- and bilayer part of the flake is not only excluded spatially, because the affected region stretches over several measurement spots to more than 8 μm away from the mono-/bilayer border, but also such a case would be reflected in a more complex 2D band region consisting of overlapping peaks from both the mono- and the bilayer. On top of that, concerning the immediate surroundings of the affected area, there is no sign of compression, only different levels of tension > 0.3%. Hence, the presence of a locally compressed region is quite improbable.

Moreover, to exclude the possibility of the $G_2$ being a spike or a ghost band, we emphasize again that its presence is always strictly limited to the marked region and is always accompanied by the same monolayer-like 2D band as presented in Fig. 4b(top), main text. This fact concerns several hundreds of spectra measured only on this flake, which have been processed both automatically and manually to ensure that the presence/absence of such a band would not be overlooked. Finally, it is well documented that the $E_u$ mode can also be activated by injecting different carrier concentrations in the two layers.[2-5] Such a scenario is not possible in our case, since the bilayer graphene is sandwiched between two insulating polymer layers.